\documentclass{article}

     \usepackage[preprint, nonatbib]{neurips_2019}

\usepackage[utf8]{inputenc} 
\usepackage[T1]{fontenc}    
\usepackage{hyperref}       
\usepackage{url}            
\usepackage{booktabs}       
\usepackage{amsfonts}       
\usepackage{nicefrac}       
\usepackage{microtype}      
\usepackage[pdftex]{graphicx}
\usepackage{multirow}

\title{Genome Variant Calling with a Deep Averaging Network}

\author{
  Nikolai Yakovenko \\
  NVIDIA \\
  \texttt{nickyakovenko@gmail.com} \\
   \And
  Avantika Lal \\
  NVIDIA \\\
  \texttt{alal@nvidia.com} \\
   \And
  Johnny Israeli \\
  NVIDIA \\\
  \texttt{jisraeli@nvidia.com} \\
   \And
  Bryan Catanzaro \\
  NVIDIA \\\
  \texttt{bcatanzaro@nvidia.com} \\
}

\begin{document}

\maketitle

\begin{abstract}

Variant calling, the problem of estimating whether a position in a DNA sequence differs from a reference sequence, given noisy, redundant, overlapping short sequences that cover that position, is fundamental to genomics. We propose a deep averaging network designed specifically for variant calling. Our model takes into account the independence of each short input read sequence by transforming individual reads through a series of convolutional layers, limiting the communication between individual reads to averaging and concatenating operations. Training and testing on the precisionFDA Truth Challenge (pFDA), we match state of the art overall 99.89 F1 score. Genome datasets exhibit extreme skew between easy examples and those on the decision boundary. We take advantage of this property to converge models at 5x the speed of standard epoch-based training by skipping easy examples during training. To facilitate future work, we release our code, trained models and pre-processed public domain datasets\footnote{\url{https://github.com/clara-genomics/DL4VC}}.

\end{abstract}

\section{Introduction}
Genome variant calling is an important problem in computational biology. Distinguishing between candidate variants and the reference genome forms a core input into most downstream genomic studies. The uses range from cancer risk prediction to ancestry studies. A typical human genome contains 3.4 million known short variants (less than 50 basepairs) in trusted regions alone. Small changes in DNA can have large impacts on biological traits. Even one SNP (single nucleotide polymorphism) can have a decisive effect on a downstream classification. Thus, in order for a variant calling system to be useful, it must provide recall and accuracy of over 99\%. 

Introduced on the precisionFDA (pFDA) Truth Challenge, DeepVariant~\cite{poplin2018universal} demonstrated that deep neural networks can be competitive with traditional variant calling methods. More recent DeepVariant versions have outpaced state of the art non-deep learning variant calling tools such as GATK (Genome Analysis Toolkit)~\cite{mckenna2010genome} and Sentieon~\cite{freed2017sentieon} on several human genome benchmarks. They also showed that their network adapts to new modalities such as instrument changes, given enough high quality training data ~\cite{DVReadshift, DVPacBio}. 

However, DeepVariant adapts the Inception network~\cite{Szegedy_2016_CVPR} that was designed for image classification. Training and inference therefore requires transforming the genomic input data into 300x300 pixel RGB images. This motivates investigation into whether a deep learning model designed directly for variant calling could do better.

We propose a custom architecture for variant calling. This model transforms individual reads through a series of convolutional layers, and limits the communication between reads to averaging and concatenating. Training and testing on pFDA, we match state of the art overall F1 score. Genome datasets exhibit an extreme skew between easy examples and those on the decision boundary. We take advantage of this property to converge models at 5x the speed of standard epoch-based training.

\section{Background}

{\bf Human genome}
The human genome consists of 3.2 billion base pairs (each base is one of adenine (A), cytosine (C), guanine (G), and thymine (T)), split across 23 chromosomes. Individuals differ from a “reference human genome” in approximately 1/1000th of those locations\footnote{in trusted regions, ignoring structural variants}. These 3-4 million differences are known as variants, of which there are three major types:
\begin{itemize}
    \item SNP (single nucleotide polymorphism) -- a single base replacement. Denoted {A -> T}
    
    \item Insertion -- one or more bases are added at a reference location. Denoted {A -> ATT}
    
    \item Deletion -- one or more bases are removed at a location. Denoted {ATT -> A}
\end{itemize}

Inserts and deletion are referred to jointly as “Indels.” Within a human genome, SNPs outnumber Indels approximately 10-1. Indels are more difficult than SNPs to classify properly, and thus classification accuracy on SNPs and Indels is usually reported separately. A human genome is present in two copies, with one copy inherited from each parent. Thus SNP and Indel variants are sub-classified into two types:
\begin{itemize}
    \item Homozygous -- the same variant occurs in both copies of the genome.
    \item Heterozygous -- a variant occurs in one copy but not the other.
\end{itemize}
Approximately two thirds of variants in a human genome are heterozygous.

There are also “multi-allele” variants, where a different variant occurs on each strand of the DNA, in a given reference location. Multi-allelic variant sites are rare but not insignificant. There are approximately 30,000 such locations, out of 3-4 million variants, about 1\% of the data. See Table \ref{table:long-insert-example} for example of a complex multi-allelic site.

There are several versions of the reference human genome. The precisionFDA Truth Challenge is based on the hs37d5 standard, while most recent work is done with the updated hg38 version of the reference. 


{\bf Single read alignments}
Sequencing a human genome starts with collecting short “reads” of sequenced DNA fragments. These reads are typically less than 300 bases, depending on the sequencing machine used \footnote{Most short read sequencing takes place on Illumina machines, HiSeq (older) and NovaSeq (post year 2017).}. 
These single reads are aligned to the reference genome, using partial string-matching algorithms \cite{li2009fast}. 
This alignment process works reasonably well in most locations of the genome, although string matching can lead to indeterminate results, within long repeat regions of the genome \cite{li2009fast}. 

\begin{figure}[!ht] 
  \begin{center}
    \includegraphics[width=0.9\linewidth]{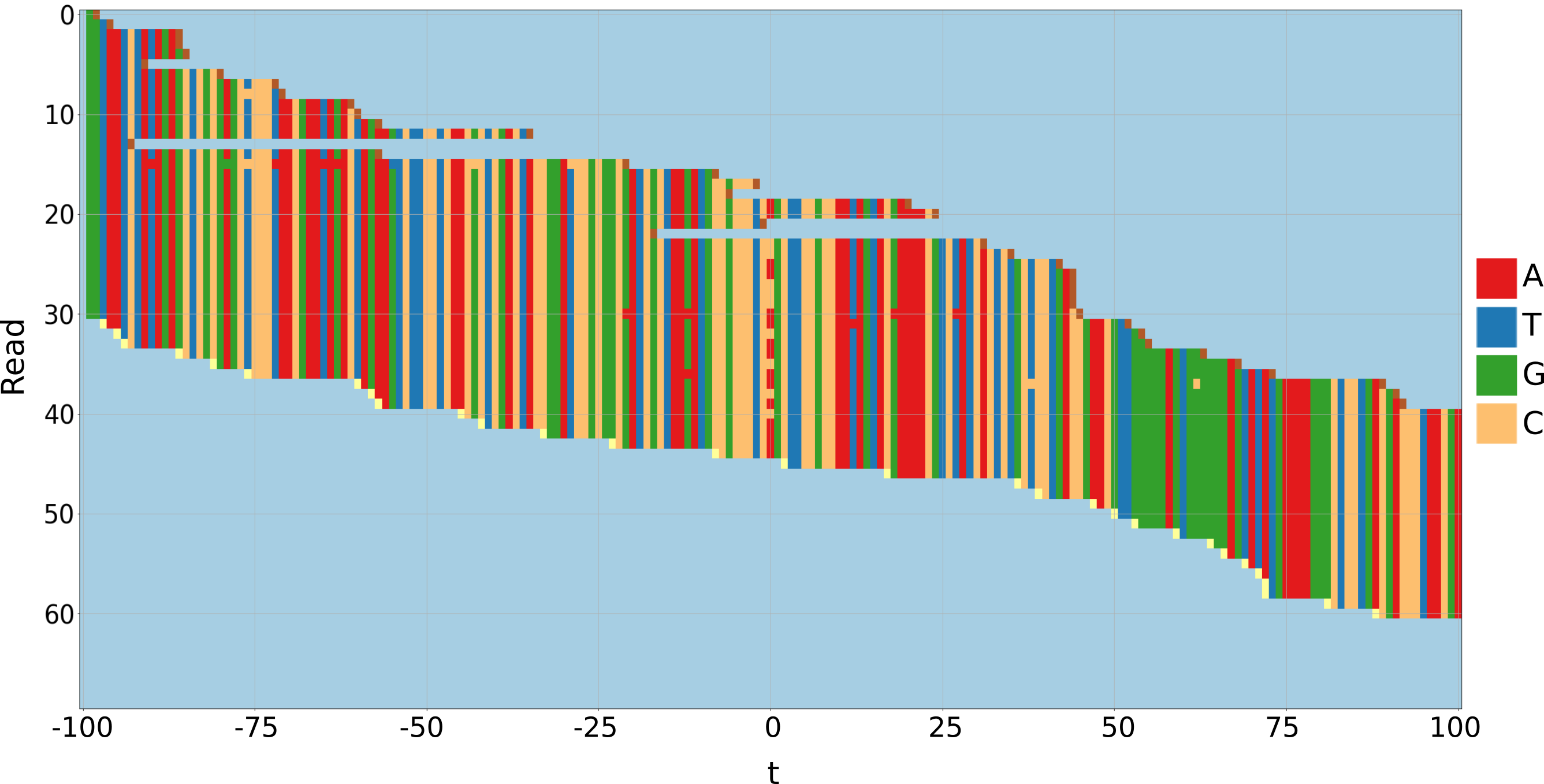}
    \caption{Each row in the pileup represents an independent sequencing read. The time axis shows genomic position, centered at a variant candidate, in this case, a heterozygous SNP.}
    \label{fig:pileup}
  \end{center}
\end{figure}

{\bf Calling variants}
Variant calling is the process of calling variants -- creating the diff between the reference genome and a newly sequenced genome -- based on information from a “pileup” of aligned short reads. This process typically proceeds at a high level as follows. First, we align the short reads to the reference genome. Second, we generate candidate variants -- a high recall, low precision set of (almost all) possible variants. Third, we score variant probabilities based on local information around the variant in question, such as a reads pileup as demonstrated in Figure \ref{fig:pileup}. This work focuses on the third step - we use traditional techniques to align reads and generate candidate variants.


The presence of even a single variant can lead to the diagnosis of an inherited disease, thus the aim is to classify all variants with a high degree of accuracy. Human genomes are usually sequenced with enough coverage depth to allow variant calling algorithms to achieve 99\% precision and recall overall. Accuracy is much lower for challenging regions such as long Indels and repeat regions.

\begin{table*}[!ht]
\centering
\caption{Examples of a multi-allele location, which is classified correctly by our model.} 
\label{table:long-insert-example}
\makebox{}{
\resizebox{0.8\textwidth}{!}{%
\begin{tabular}{ccc|cl|cc|c}
\toprule
Genome & Chrom & Location & Ref & Variant & Depth & Allele Frequency & Truth \\
\midrule
HG002 & 6 & 51564718 & A & AGT & 33 & 0.121212 & False \\
HG002 & 6 & 51564718  & A &  AGTGC  & 33 & 0.030303  & False \\
HG002 & 6 & 51564718  & A &  AGTGT  &  33 & 0.151515 & True \\
HG002 & 6 & 51564718 & A  & AGTGTGT & 33 & 0.303030 & True \\
\bottomrule
\end{tabular}%
}
}
\end{table*}

\section{Related Work} 

GATK\cite{mckenna2010genome}, the most widely used variant calling tool, uses a combination of logistic regression, hidden Markov models, and naive Bayes, combined with hand-crafted features to remove likely false positives. 

DeepVariant~\cite{poplin2018universal} demonstrated that a deep neural network trained with gradient descent could produce variant calls competitive with statistically based state of the art methods.

The DeepVariant method involves converting  aligned sequence reads for each candidate variant region into an RGB image, along with additional read information, such as the base quality scores. This image is fed into the Inception image classification convolutional neural network, predicting a softmax over three classes for each candidate variant: \{no variant (false positive), heterozygous variant, homozygous\}.

After the pFDA result, DeepVariant significantly improved their model, for both SNPs and Indels (see Table \ref{table:pFDA-results}), by training on 10x additional human genomes. This demonstrates that the deep neural network approach benefits from additional training data, and would likely out-pace statistically driven and hand-tuned approaches to variant calling, given enough quality training examples. Although additional data helps, the additional data is not public, and so for reproducibility, in this work we focus on approaches trained on the pFDA dataset. 

The DeepVariant method has since been applied to variant calling for non-human genomes \cite{wang2018genomic, DVrice} , as well as to the output of other sequencing machines such as Illumina NovaSeq~\cite{DVReadshift} and technologies, such as PacBio Circular Consensus Sequencing~\cite{DVPacBio}.

\subsection{Differences with DeepVariant}

We propose a new deep neural network for variant calling. 

The task is one of counting and comparing single reads to form a consensus, in this case for the likelihood of a heterozygous or homozygous variant. The individual reads are more like a sequence of letters or symbols than an image. Yet recent attempts to represent variant calling as sequences and not images have not been competitive with DeepVariant, or other state of the art methods \cite{luo2019multi,torracinta2016training}.

Our goal is simple: design a deep learning network that processes individual reads independently, unlike DeepVariant’s 2-dimensional convolutional operations. Information between different reads must ultimately be shared to produce the result. Our aim was to do so in a small number of simple operations, specifically as average pooling across all reads in a pileup. 
By doing so, we take advantage of the structure of the data: since the reads are each produced independently, we hypothesize that a neural network that processes the reads independently more accurately reflects the structure of the problem.

\section{Experiments}
\subsection{precisionFDA Truth Challenge}
The precisionFDA Truth Challenge, sponsored by the FDA in 2016, is a competition on genomics data. Teams compete to predict variants on a genome dataset for HG002 (human genome 002 from Genome in a Bottle -- GIAB), with training provided for HG001 (human genome 001, also from GIAB). Both training and test set BAMs are built from reads from an Illumina HiSeq2500 machine, downsampled to 50x coverage. Within high confidence trust regions (known for HG001, unknown but similar for HG002) there are approximately 3.4 million true variants. 

Teams were measured on their accuracy (F1 score) for predicting variants on SNPs and Indels, with prizes awarded for the highest precision, recall and F1 for SNPs and Indel variants. The top results are reported on the precisionFDA website\footnote{\url{https://precision.fda.gov/challenges/truth}}, 
and reproduced in Table \ref{table:pFDA-results}.

Teams are expected to predict the zygosity of variants, as well as joining any multi-allele sites. While accuracies for zygosity and multi-allele are not reported for the challenge, predicting either of these categories wrong results in multiple errors. We reproduced precisionFDA results for SNPs and Indels by running the Hap.py program \cite{happy} on the HG002 variant calls.


\subsection{Training with additional data} 
DeepVariant reports their pFDA results (which won the top prize for SNPs F1), as well as a “live GitHub” version of DeepVariant, which gets the top F1 for SNPs and Indels on pFDA. This updated model was trained with 10x new HG001 datasets, demonstrating that DeepVariant’s model generalized better with more training data.

Similarly, we trained our model with the pFDA training data, and additional genome datasets, also HiSeq sequenced HG001, drawn from a public dataset \cite{telenti2016deep}. 

We demonstrate that our model also improves on the pFDA challenge when training with three additional HG001 datasets.

\section{Methods}

\begin{figure}[t!] 
  \begin{center}
    \includegraphics[width=0.9\linewidth]{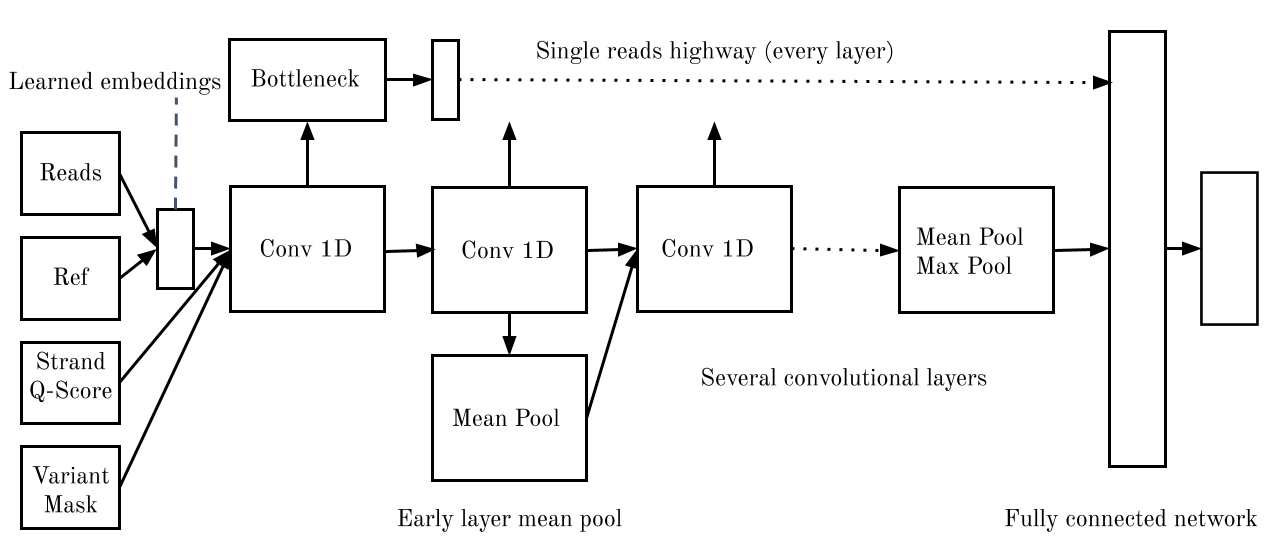}
    \caption{Network layout.}
    \label{fig:network}
  \end{center}
\end{figure}


\begin{table*}
\centering
\caption{Example read encoding for variant proposal A -> ATT.}
\label{table:var-base-encoding-example}
\makebox{}{
\resizebox{0.7\textwidth}{!}{%
\begin{tabular}{l|ccccc|ccccc}
\toprule

Read Bases & G & A & T & T & C & G & A & - & - & C  \\
Reference Bases & G & A & - & - & C & G & A & - & - & C\\
Base Quality & 70 & 60 & 50 & 45 & 50 & 60 & 50 & 65 & 35 & 55\\
Strand Direction & 1 & 1 & 1 & 1 & 1 & 2 & 2 & 2 & 2 & 2 \\
Reference Mask & 0 & 0 & 0 & 0 & 0 & 0 & 1 & 1 & 1 & 0 \\
Variant Mask & 0 & 1 & 1 & 1 & 0 & 0 & 0 & 0 & 0 & 0 \\
Var Length Mask & 0 & 1 & 1 & 1 & 0 & 0 & 1 & 1 & 1 & 0 \\

\bottomrule
\end{tabular}%
}
}
\end{table*}


\subsection{Network}
Our model transforms individual reads through a series of 1-dimensional convolutions, pools the final layer outputs across all reads, and outputs final predictions through a fully connected neural network, as illustrated in Figure \ref{fig:network}.

{\bf Encoding individual reads}
The input consists of a pileup of aligned reads, such as in Fig.\ref{fig:pileup}, and a variant candidate proposal. In addition, the network takes in base quality scores, strand direction for each read, and masks representing the reference and the variant proposal, as shown in Table \ref{table:var-base-encoding-example}.



\begin{table*}
\centering
\caption{Model details.}
\label{table:model-details}
\makebox{}{
\resizebox{0.75\textwidth}{!}{%
\begin{tabular}{l|cc}
\toprule
Category & Parameters & Values \\
\midrule
Pileup & maximum single reads & 100  \\
       & read length & 201 (100 to left and right) \\
\midrule
Input & embedding dimensions & 20 \\
 & total input dimensions & 100x201x45 \\
\midrule
Conv layers & number of  layers & 7 \\
            & residual layers & 5,6,7 \\
            & output channels & 128 \\
            & activation & ReLU \\
            & batch normalization \cite{BatchNormIoffe} & true \\
            & dilation~\cite{DilatedConvYu} & 2, except first layer \\
\midrule
Pooling & mean pool, max pool & final layer \\
        & early mean pool & after layer 2 \\
\midrule
Highway & reduce dim to 32 channels & every layer \\
    & final highway output & 100x32 per layer \\
\midrule
FCN & input dimensions & 73856 \\
    & layers & 1025, 256 \\
    & activation & ReLU \\
    & dropout & 0.1 \\
\midrule
Training & optimizer & ADAM \\
 & learning rate & 0.0002 \\
 & focal loss & $\gamma=0.2$ \\
 & label smoothing & $\epsilon = 0.001$ \\
 & easy example window & $2\epsilon$ \\
 & easy examples skip rate & 0.85 \\

\bottomrule
\end{tabular}%
}
}
\end{table*}

The reads and reference bases are expanded into a learned multi-dimension embedding, similar to learned embeddings for a deep language model \cite{Mikolov2013}. We also add sinusoidal positional embeddings to each dimension of the learned base embeddings, as introduced in \cite{Transformer2017}.



{\bf One-dimensional convolutional layers}
We transform the individual reads through as a series of convolutional layers, with small one-dimensional convolutional filters, not sharing information between single reads.


{\bf Final layer pooling}
We combine disparate single reads by performing mean pooling and max pooling operations across all locations and channels. The mean and max pool outputs are then flattened, and input to a fully connected network.

This network, similar to the DAN (Deep Averaging Network) \cite{Iyyer-DAN} has the additional property of ignoring read order in the pileup, since all operations are performed at the individual read level, then the final outputs are averaged across all reads. 

{\bf Highway layers}
Passing all read level information through seven convolutional layers followed by a wide pooling layer may not be efficient. 
We concatenate a small amount of information, for every read, directly to the final fully connected network. Details in Table \ref{table:model-details}.  


{\bf Fully connected network for variant candidate classification}
We connect the concatenated outputs of the pooling layers and the highway layers, to a fully connected network, including dropout and ReLU activation layers after every fully connected layer. The final output is a softmax prediction for \{no variant, heterozygous, homozygous variant\}.


{\bf Additional Early Pooling layer}
Notably, information between disparate reads is not shared until the final layer pooling. To allow read comparison computation to take place in the convolutional layers instead of in the fully connected network, we insert a second mean pooling layer after the second convolutional layer. 

\subsection{Application considerations}
Although the main focus of this work is the variant calling neural network, we describe the other components necessary for this network to function as a complete variant calling system.

{\bf Candidate generation}
The goal of candidate generation is to produce a high recall set of candidate variants that we will then be scored by our variant calling network. We use a simple heuristic to generate candidates. First, we count any mapped reads that disagree with the reference at any location in the trusted regions. Then, we create a variant candidate at any location, as long as the allele frequency (percentage of reads matching the variant candidate) is above a threshold we set for high-recall.

We use thresholds of 0.05 for SNP candidates, and 0.02 for Indel candidates. For a 50x coverage dataset, this means that we accept all possible Indel variants as candidates, but we restrict very low frequency SNP candidates from our candidate dataset.

On the pFDA HG002 test set, this produces 13.4 million SNP candidates and 1.22 million Indel candidates. Our candidate generator has 99.995\% recall for SNP variants and 99.48\% recall for Indel variants on the HG002 test set\footnote{Our candidate generator misses 124 SNPs, 316 insertions and 1433 deletions within the HG002 trusted region. Since our neural network scores candidates but does not propose them, our overall accuracy depends on good candidate generation, and we believe a more sophisticated candidate generation procedure would further improve accuracy. In other words, candidate generation bounds the accuracy of our model, as shown in Table \ref{table:pFDA-results} }.




{\bf Thresholding} 
Our model produces softmax outputs \{no variant, heterozygous, homozygous variant\} for each candidate. To produce actual variant calls, we need to threshold both variant truth, and zygosity. 

Using default thresholds of 0.3 for variant calling and 0.5 for zygosity produces results that are close to those with optimal thresholding. Ideally, we would use a small thresholding dataset, separate from the training and test set.



{\bf Multi-allele inference}
We take a naive approach to multi-allele training and inference. All alleles, are trained and inferred as independent examples. We merge the top two alleles, unless the top allele is homozygous and the second allele is below a 0.95 variant probability.
%

Following this simple rule, we classify multi-allelic sites on the pFDA test set with 0.98154 F1. 

\subsection{Training} 
The genome variant calling dataset is heavily skewed, not just by label frequency, but by the difficulty of the training examples. 

We train our model with label smoothing \cite{LabelSmoothing} to avoid saturating the softmax outputs. 
We also found that focal loss \cite{Lin2017FocalLF} helps convergence. Focal loss reduces the loss weight on well-classified examples, increasing gradient contibutions from mis-classified examples. 

After one epoch of training, 99.08\% of training examples have been classified correctly, within $2.0*\epsilon$ of the true label, where $\epsilon$ is the label smoothing value (after two epochs, easy examples grow to 99.70\%). With focal loss, we are already driving the loss weight on those examples to zero, thus it would save us a lot of training time just to skip those examples.
We are not aware of similar techniques of active data downsampling for skewed supervised learning tasks, although similar techniques are widely used in reinforcement learning \cite{Mnih2013PlayingAW}.

Our pFDA training starts with 14,656,643 training candidate examples, 4 epochs of training, no decay and 300 global batch size. 
We trained our model in the PyTorch~\cite{PyTorch} framework, on a single NVIDIA Tesla V100 GPU. Additional training details are listed in Table \ref{table:model-details}.

\begin{table*}[!ht]
\centering
\caption{pFDA Truth Challenge results and   results with supplementary training data.}
\label{table:pFDA-results}
\makebox{}{
\resizebox{0.90\textwidth}{!}{%
\begin{tabular}{cc|ccc|ccc}
\toprule
& Type & F1 & Recall & Precision & TP & FN & FP \\
\midrule
rpoplin-dv42 & Overall & 0.998597 &	0.998275 &	0.998919 &	3,393,136 &	5,864 &	3,671\\
(DeepVariant) & Indels & 0.989802 &	0.987883 &	0.991728 &	340,370 &	4,175 &	2,839\\
 & SNPs & 0.999587 &	0.999447 & \textbf{0.999728} &	3,052,766 &	1,689 &	832\\
 \midrule
dgrover-gatk & Overall &	0.998905 &	0.999005 &	0.998804 &	3,395,497 &	3,381 &	4,066\\
(GATK)	& Indels & \textbf{0.994008} & 0.993455 &	\textbf{0.994561} &	342,154 &	2,254 &	1,871\\
	& SNPs &	0.999456 &	0.999631 &	0.999282 &	3,053,343 &	1,127 &	2,195\\
\midrule
astatham-gatk &	Overall &	0.995679 &	0.992122 &	0.999261 &	3,372,103 &	26,775 &	2,493\\
(GATK)	& Indels &	0.993422 &	0.992401 &	0.994446 &	341,788 &	2,617 &	1,909\\
	& SNPs &	0.995934 &	0.992091 & \textbf{0.999807} &	3,030,315 &	24,158 &	584\\
\midrule
bgallagher-sentieon &	Overall &	0.998626 &	0.998910 &	0.998342 &	3,395,174 &	3,706 &	5,638\\
(Sentieon)	& Indels &	0.992676 &	0.992140 &	0.993213 &	341,703 &	2,707 &	2,335\\
	& SNPs &	0.999296 &	\textbf{0.999673} &	0.998919 &	3,053,471 &	999 &	3,303\\
\midrule


       
Ours & Overall & \textbf{0.998924} & \textbf{0.999076} & 0.998772  &  3,394,460 & 4,172 & 3,138 \\
(pFDA) & Indel & 0.992949 & \textbf{0.994708} & 0.991196 & 340,802 & 3,027 & 1,813 \\
       & SNPs & \textbf{0.999596}   & 0.999566 & 0.999625 & 3053658 & 1,145 & 1,325 \\ 
       
\midrule
\midrule
DeepVariant* & Overall & 0.99932	& 0.99909	& 0.99955	& 3,412,193	& 3,104	& 1,548\\
(V0.4) & Indel & 0.99507 &	0.99347 &	0.99666 &	357,641 &	2,350	& 1,198\\
(+10 genomes) & SNPs & 0.99982	& 0.99975	& 0.99989	& 3,054,552	& 754	& 350\\
 
\midrule
 Ours & Overall & 0.999139 &	0.998874 &	0.999404 &	3,394,796 &	3,827 &	2,023\\
(+3 genomes) & Indel & 0.994469 &	0.992227 &	0.996722 &	341,195 &	2,673 &	1,122\\
 & SNPs & 0.999664 &	0.999622 &	0.999705 &	3,053,601 &	1,154 &	901\\

\bottomrule
\end{tabular}%
}
}
\end{table*}

\section{Results}

When training on the pFDA HG001 dataset, our model achieves a better overall F1 than DeepVariant “pFDA,” which was limited to the pFDA dataset. Our model matches the best overall F1 when combining SNPs and Indels, and it would have a prize for the best Indel recall, according the rules of the pFDA Truth Challenge.

When trained with three additional HiSeq HG001 datasets, our model achieves a better result on SNPs, and also on Indels, than any submission to the pFDA challenge. This result also closes the gap between our pFDA submissions and the DeepVariant v0.4 result, which was trained with 10x additional datasets. 

Thus we demonstrate that our model, like DeepVariant, benefits from more training data, even when that data is a different run of a similar sequencing machine, on the same underlying genome. These models generalize better to the pFDA HG002 genome, suggesting the gains are not simply overfit to the HG001 Truth Set.\footnote{There is a discrepancy between DeepVariant v0.4 results \cite{poplin2018universal} and official pFDA results. An additional 15,000 Indels have been added to HG002 evaluation, despite reference to v3.2.2 of the GIAB Truth Set.}

\subsection{Ablation studies}
The details of our neural network architecture are described in Table \ref{table:model-details}. In Table \ref{table:ablation-results}, we demonstrate some ablation studies, from reducing the number of layers, to removing network components such as the highway layers. 

Specifically we notice that the model generalized less well, when the highway layers are removed. When the pooling layer dimensions are reduced from 128 to 64 or 32 channels, this greatly reduced the model’s parameter count, and also increases the test loss and decreases test accuracy. However, a smaller highway dimension appears optimal for a smaller pooling channel output, suggesting that these parameters must be kept in balance.

We also notice that down-sampling easy examples after the first epoch, appears to improve test accuracy, as well as save 5x in training time.

Lastly, we notice that test results are slightly unstable. This effect is greatly diminished when training pFDA with additional datasets. This not only improves generalization as show in Table \ref{table:pFDA-results}, but the model appears to be more stable when increasing the number of difficult examples by training on several genomic datasets. 



\begin{table*}[!ht]
\centering
\caption{Ablation studies, when changing training configurations from Table \ref{table:model-details}. Since every variant is important, notice the difference in the FN and FP counts, as well as overall F1 scores.}
\label{table:ablation-results}
\makebox{}{
\resizebox{0.90\textwidth}{!}{%
\begin{tabular}{c|ccc|ccc}
\toprule
 &  TP & FN & FP &   F1 & Recall & Precision \\
\midrule
(baseline) &  3,394,460 & 4,172 & 3,138 &    0.998924  &  0.999076  &  0.998772 \\
\midrule
no early pool & 3,394,352 & 4,264 & 3,199  &  0.998902  &  0.999058  &  0.998745 \\
0.01 label smoothing & 3,394,098 & 4,522 & 2,984 &   0.998895  &  0.999122  &  0.998669  \\
0.1 label smoothing & 3,392,884 & 5,759 & 4,925 &   0.998428  &  0.998551  &  0.998306 \\

no Strand, no Q-Scores & 3,392,780 & 5,858 & 4,369 &   0.998495  &  0.998714  &  0.998276 \\

no HW layers & 3,393,121 & 5,513 & 4,297 & 0.998557  &  0.998735  &  0.998378  \\
128 -> 64 final channels & 3,394,159 & 4,457 & 3,204 &  0.998873  &  0.999057  &  0.998689 \\
128 -> 32 final channels  & 3,393,643 & 4,991  & 3,335 &  0.998775  &  0.999018  &  0.998531 \\
6 conv layers  & 3,393,907 & 4,710 & 3,142 &   0.998845  &  0.999075  &  0.998614  \\
5 conv layers  & 3,394,147 & 4,486 & 2,861 &  0.998919  &  0.999158  &  0.998680  \\
4 conv layers  & 3,393,100 & 5,514 & 3,414 &  0.998686  &  0.998995  &  0.998378  \\
no focal loss & 3,393,591 & 5,041 & 3,521 &  0.998740  &  0.998964  &  0.998517  \\


\bottomrule
\end{tabular}%
}
}
\end{table*}



\section{Conclusion}
We presented a deep neural network for genome variant calling. Our work shows that it is possible to solve the variant calling problem with a individual read-sequence level model, without any two-dimensional convolutions or pooling. 

Our approach generalizes well enough to match state of the art on the pFDA Truth Challenge, and it benefits substantially from additional training data. We also demonstrate how it is possible to converge a variant calling model more quickly, by aggressively down-sampling training on well-classified examples. 

We believe this is a useful first step toward genomics-specific neural network architectures. We hope to see others build on top of this approach in the years to come. 

\section{Acknowledgements}
Thanks to Mike Vella, Joyjit Daw, Michelle Gill and the NVIDIA genomics group for help with genomics tooling, as the variant calling problem was new to us before embarking on this work. Thanks also to Andrew Tao, Patrick LeGresley, Boris Ginsburg, Robert Pottorff, Ryan Prenger and Saad Godil of NVIDIA's applied deep learning research group, for insightful discussions about the neural network design and training of our model. Our methods borrow from disparate deep learning problems of speech generation, NLP, computer vision and reinforcement learning. Finally thanks to Jason Chin, Yih-Chii Hwang and the DNANexus research team, as well as Ali Torkamani and the Scripps Research Translational Institute for helping us with additional sources of human genome data in the public domain, beyond the precisionFDA Truth Challenge. 


\bibliographystyle{ieee}
\bibliography{aaai}




\end{document}